\documentclass[12pt]{article}

\usepackage{epsf}
\begin{document}

\bigskip
\bigskip
\bigskip
\begin{center}
{\bf \large ANALYSIS AND INTERPRETATION}
{\bf \large OF X-RAY PROPERTIES}\\
{\bf \large OF BLACK HOLE CANDIDATES}
\end{center}
\vskip 2.5cm
\begin{center}
{\bf Sergey P. Trudolyubov}\\
\vskip 2.5cm
{\bf A thesis submitted in partial fulfillment of}\\
{\bf the requirements for the degree of}\\
{\bf DOCTOR OF PHILOSOPHY (Astrophysics)}\\
{\bf at the}\\
{\bf SPACE RESEARCH INSTITUTE}\\
{\bf RUSSIAN ACADEMY OF SCIENCES}\\
{\bf December 1999}\\
\vskip 1.7cm
{\it Advisor Dr. Marat R. Gilfanov}
\end{center}

\clearpage

{\bf \hfill Summary \hfill}

\bigskip

The thesis is mainly devoted to the study of spectral and timing 
characteristics of the X-ray emission from Galactic black hole 
candidates GRS 1915+105, GX 339--4, 4U 1630--47, XTE J1748--288, 
GRS 1739--278, KS/GRS 1730--312, GRS 1737--31. The main results 
include: observation of the correlated evolution of spectral and 
timing parameters of these sources and its interpretation in the 
framework of two--phase model of the accretion flow near the 
compact object; development of the quantitative model for the 
evolution of GRS 1915+105 during flaring state providing direct 
estimate of the disk accretion rate in the system; restrictions 
on the parameters of the spatial distribution and luminosity 
function of the Galactic hard X-ray transient sources; 
application of the the bulk motion comptonization model to the 
analytic approximation of the broad--band energy spectra of 
Galactic black hole candidates in the {\em high/very high} state 
obtained with RXTE.

\bigskip
  
{\bf Full ps--version of the thesis (in russian) can be found at:}

{\em http://hea.iki.rssi.ru/~tsp/research.html}

\clearpage

The {\bf first part} of the thesis contains a brief description of
instruments aboard the GRANAT, MIR/KVANT and RXTE observatories and 
observational techniques used in the analysis.

\begin{figure}[htb]
\hbox{
\begin{minipage}{9.0cm}
\epsfxsize=9.0 cm
\epsffile{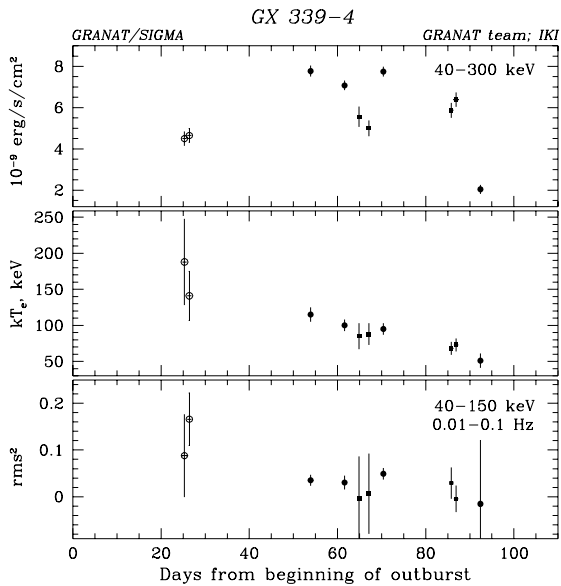}
\end{minipage}
\begin{minipage}{4.5cm}
\caption{\small The evolution of the characteristics of hard X-ray radiation 
from GX 339--4 with time since the beginning of the $1991$, $1992$ and $1994$ 
outbursts: $40 - 300$ keV energy flux -- {\em upper panel} ; 
best--fit bremsstrahlung temperature -- {\em middle panel} ; ($0.01 - 0.1$ 
Hz) fractional {\em rms}$^2$ of $40 - 150$ keV flux fluctuations -- {\em 
lower panel}. Solid circles, open circles and solid squares in each panel 
correspond to the $1991$, $1992$ and $1994$ SIGMA data respectively. 
\label{phase_general}}
\end{minipage}
}
\end{figure}

In {\bf the second part} the results of hard X-ray observations 
of well-known Galactic black hole candidate GX 339--4 with GRANAT/SIGMA 
in $1990 - 1994$ are presented. It was shown that the evolution of 
spectral and temporal characteristics of the source emission is very 
similar for individual outbursts (Fig. \ref{phase_general}). Prominent 
correlation between the hardness of energy spectrum and the level of 
aperiodic variability of hard X-ray emission of GX 339--4 has been found, 
which is typical for the black hole candidate systems in the {\em hard} 
spectral state (Fig. \ref{phase_general}). Basing on the properties of 
the long--term X-ray evolution of GX 339--4, we propose the triggering 
of the irradiation/magnetically--driven mass transfer instability in 
the low mass binary system as possible origin of the source transient 
outbursts.

{\bf The third part} of the thesis consists of 3 chapters, describing 
X-ray properties of the Galactic microquasar GRS 1915+105 in different 
spectral states.

\begin{figure}[htb]
\epsfxsize=12cm
\centerline{\epsffile{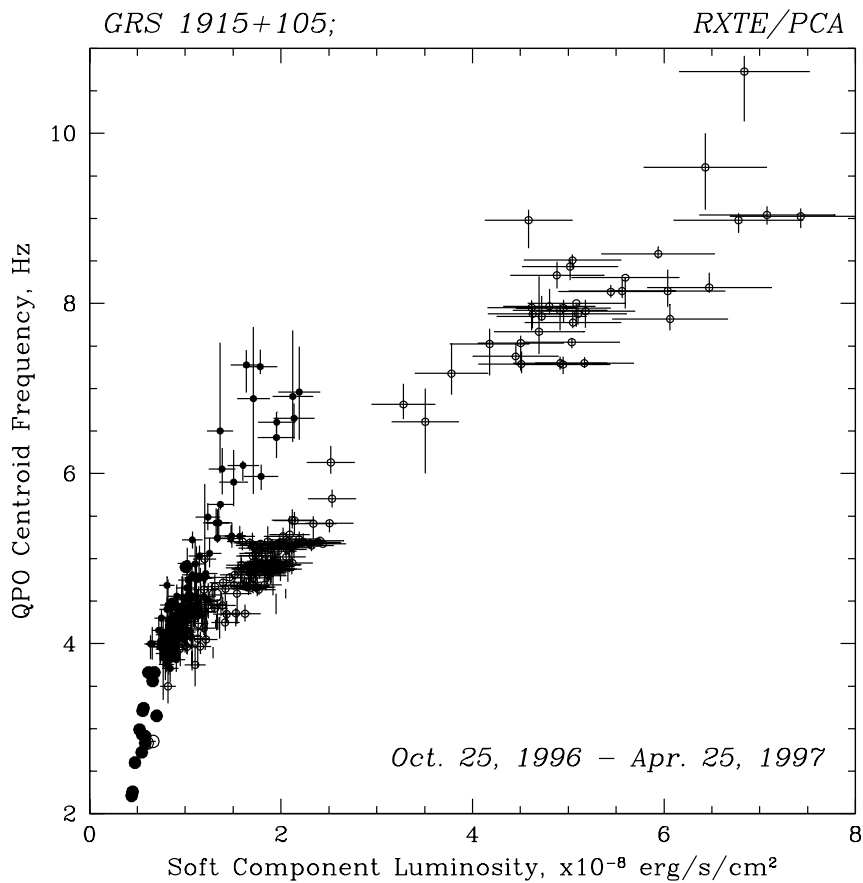}}
\caption{\small The dependence of the QPO centroid frequency on the 
estimated bolometric luminosity of the soft component obtained 
from the spectral fitting with our simplified model (multicolor 
disk black body plus power law) for the November, $1996$ -- April, 
$1997$ low luminosity state and state transitions of GRS 1915+105. 
{\it Open circles} correspond to the observations, covering 
the period of transition from the high luminosity state to the low 
luminosity state prior to November $28$, $1997$ (MJD $50415$); 
{\it filled circles} correspond to the period of low luminosity
state and following rise of the source flux. {\it Large circles} 
represent the averaging over the whole observation; the results for 
the observations with high level of variability averaged over 
$16 - 80$ {\it s} time intervals are presented by {\it small circles}. 
\label{qpo_soft_lum}}
\end{figure}

\clearpage

In {\bf\em the first chapter} we present the results of observations of 
GRS 1915+105 with PCA and HEXTE instruments aboard RXTE from October 1996 
through April 1997, when the source exhibited transitions between the 
states with high and low X-ray luminosities. With the exception of some 
individual features, the general temporal and spectral characteristics of 
the X-ray emission from GRS 1915+105 and the pattern of their evolution 
were similar to the properties of other Galactic black hole candidates in 
the so--called {\it intermediate} state, which corresponds to a transition 
between the canonical {\it low} and {\it high} states. The broad--band 
source spectrum in the $3 - 150$ keV energy range can be represented as 
a sum of a relatively weak soft component with a characteristic 
temperature of $\sim 1 - 2$ keV and a hard component; the latter can be 
fitted by a power law with a photon index of about $1.8 - 3.0$ and an 
exponential cutoff at energies $60 - 120$ keV. The temporal evolution of the 
energy spectrum is characterized by a gradual increase in its hardness 
as the total X-ray luminosity decreases from $\sim 10^{39}$ to $\sim 2
\times 10^{38}$ erg/s (for the assumed source distance of 12.5 kpc). A 
distinctive feature of the power spectrum of GRS 1915+105 during the period 
under consideration is the presence of a quasi--periodic oscillation (QPO) 
peak at frequencies $2 - 10$ Hz. The variations of the QPO centroid frequency 
are strongly correlated with the variations of spectral and timing 
parameters (in particular parameters of the soft spectral component). 
This type of correlation holds on the wide range of time scales 
(from $\sim$ seconds to $\sim$ months)(Fig. \ref{qpo_soft_lum}). Based on 
our observational data and taking into account the similarity of the 
salient features in the behavior of GRS1915+105 to the corresponding 
properties of Galactic black-hole candidates, we may assume that a further 
decrease in the source luminosity (below the observed values during the 
reported period) would result in a transition of GRS 1915+105 to a state 
close to the canonical {\it low} state. 

\begin{figure}[htb]
\hbox{
\epsfxsize=7.0cm
\epsffile{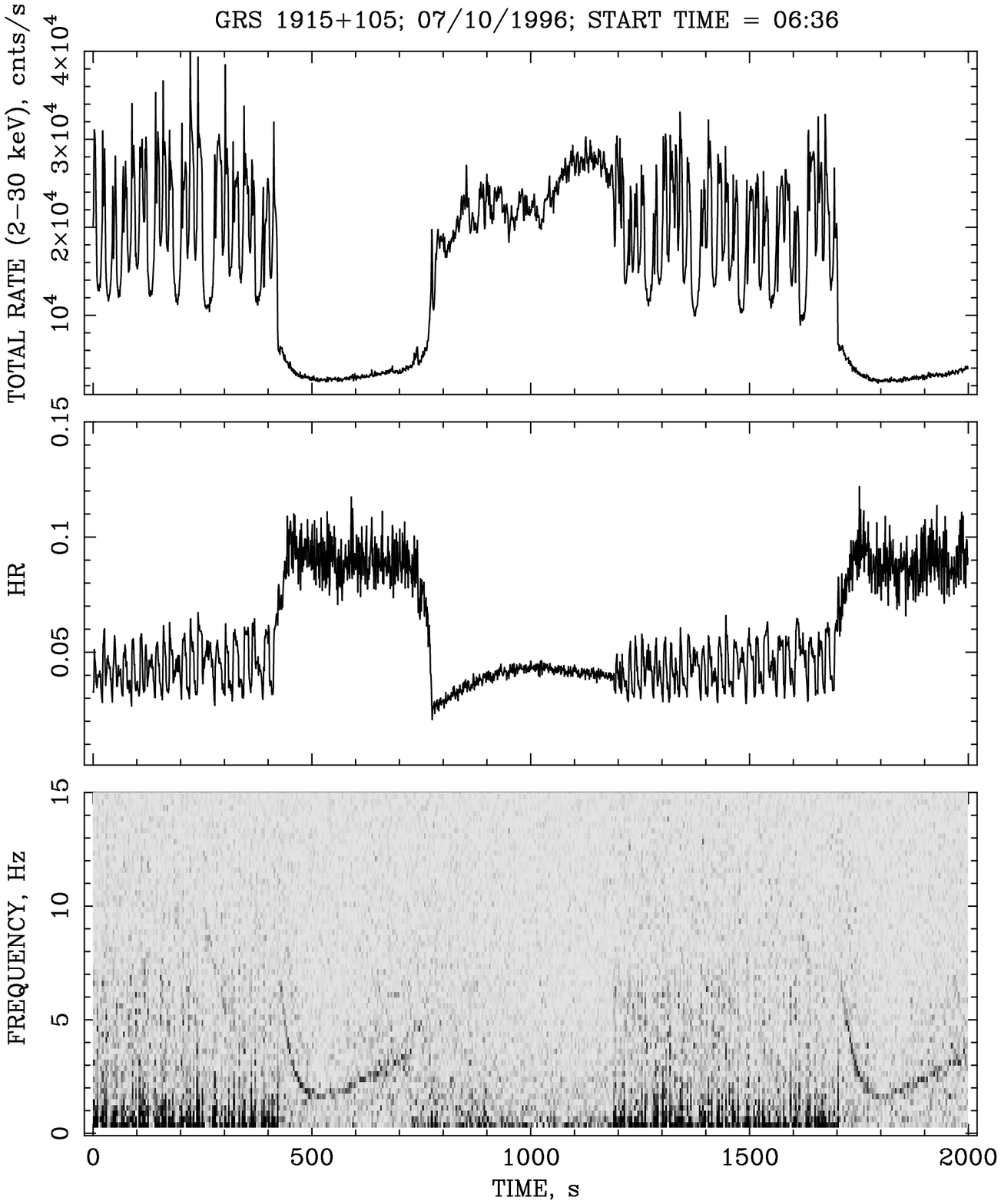}
\epsfxsize=7.0cm
\epsffile{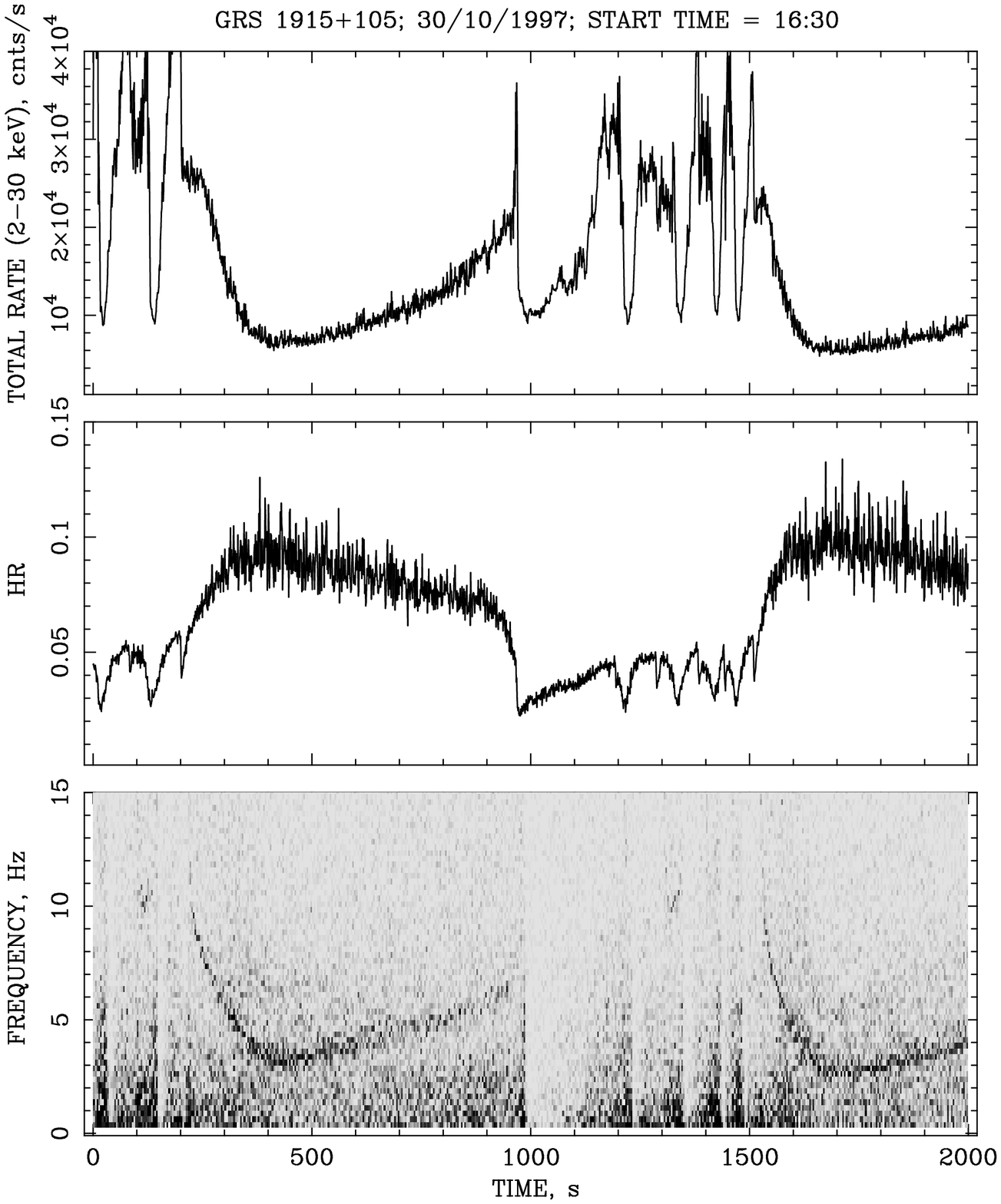}
}
\caption{\small Part of the light curves ({\em upper panels}), corresponding 
hardness ratios ($13 - 30$ keV)/($2 - 13$ keV) ({\em middle panels}) and 
dynamic PDS ({\em lower panels}) of GRS 1915+105 for the Oct. 7, 1996 ({\it 
left panels}) and Oct. 30, 1997 ({\it right panels}) observations ($2 - 30$ 
keV energy band, PCA data). The QPO peak appears as 'U'-shaped dark band in 
the {\it lower panels}. The light curves were not corrected for the
instrument dead time which was $5 - 20 \%$. The overall count rate 
corresponds to 5 Proportional Counter Units of PCA detector.
\label{grs1915_2_lc_pds}}
\end{figure}

\begin{figure}[htb]
\hbox{
\epsfxsize=6.3cm
\epsffile{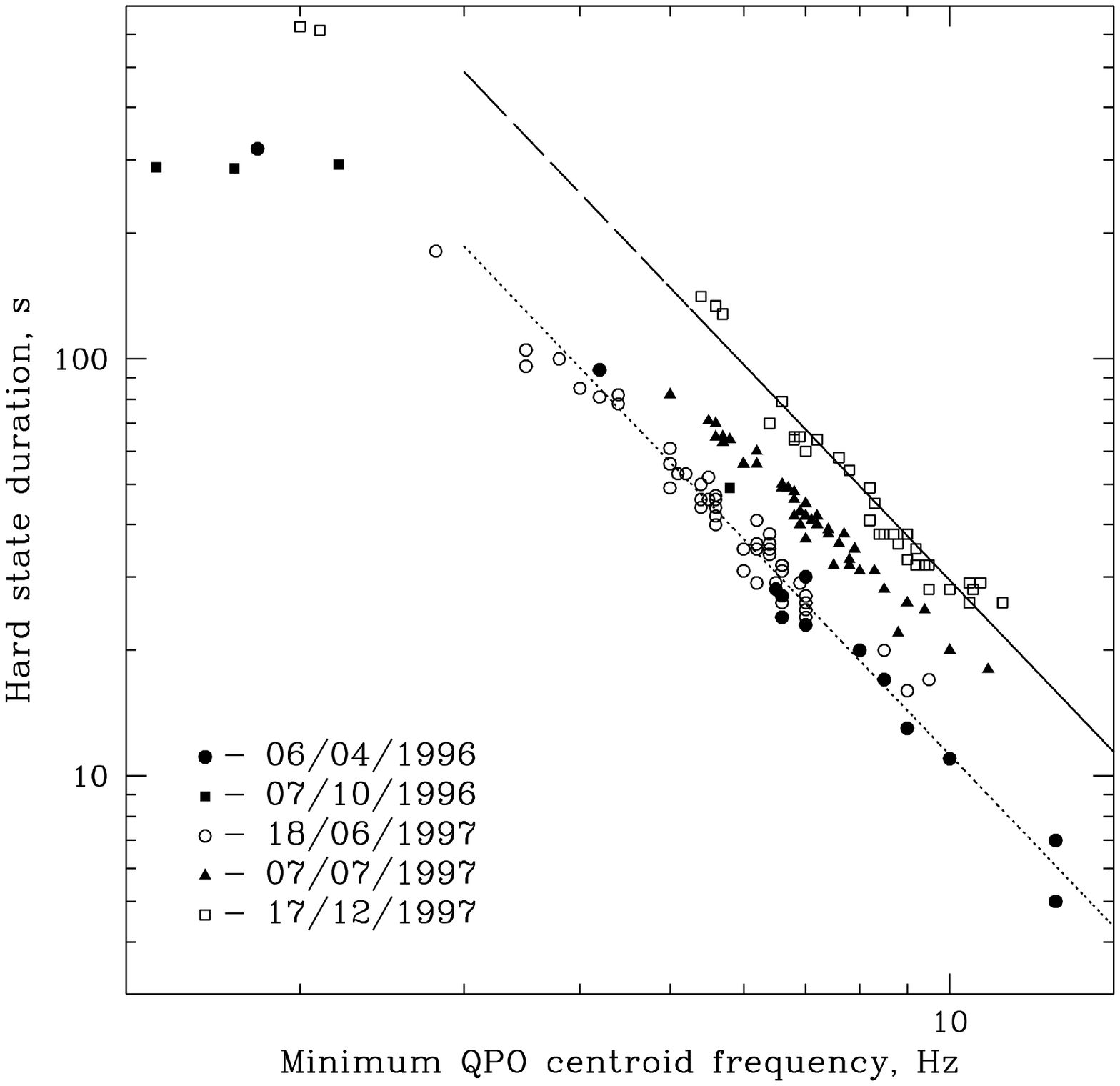}
\epsfxsize=6.3cm
\epsffile{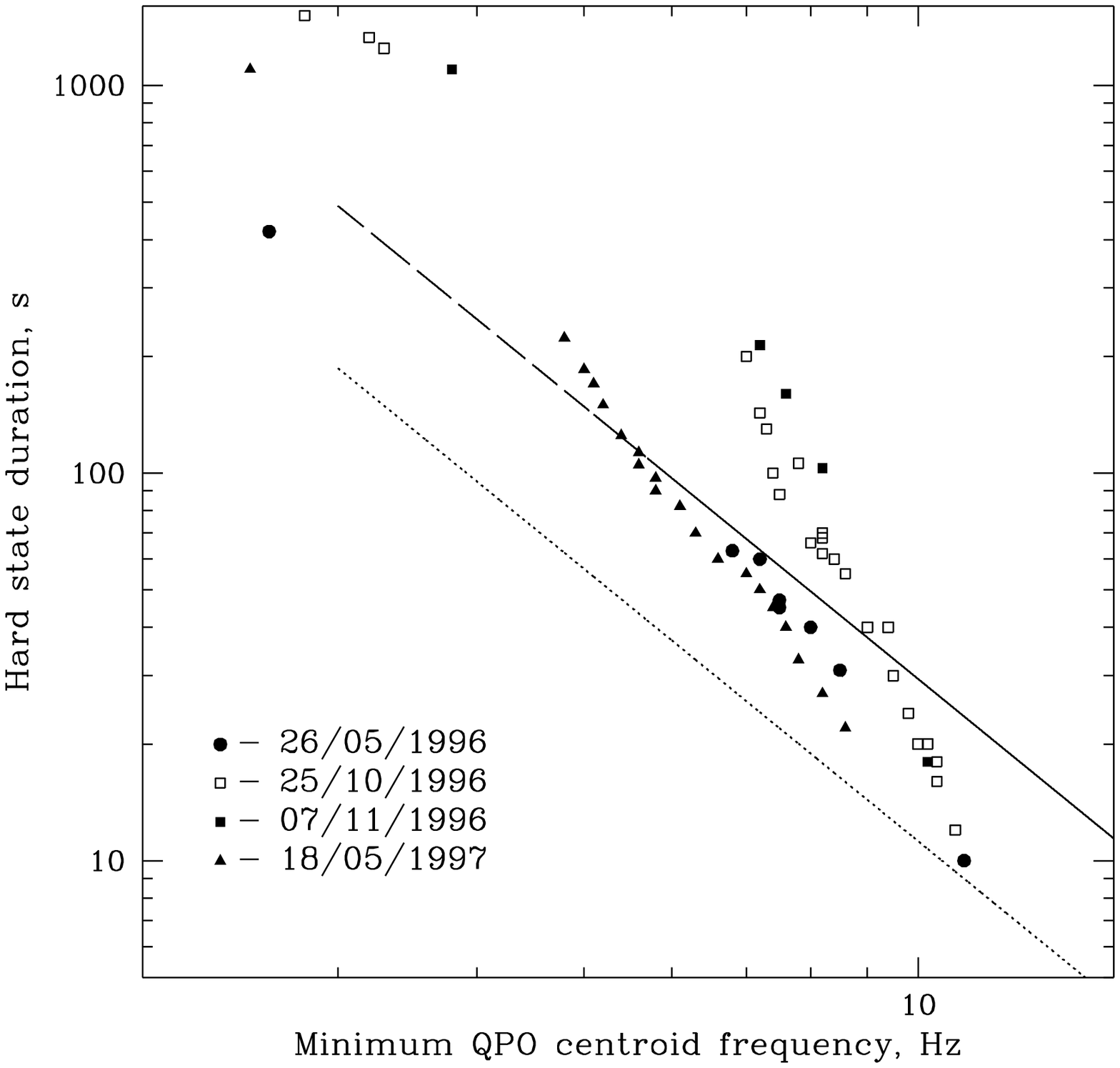}
}
\caption{\small The relation between the duration of a hard episode and a 
corresponding minimal QPO frequency for the flaring state observations 
(first group -- {\it left panel}; second group -- 
{\it right panel}). The dependences $t_{\rm visc} \propto f_{K}^{-7/3}$ of 
the viscous time scale upon the Keplerian frequency at the inner edge 
of the radiation pressure dominated disk with the mass accretion rate 
$\dot{m} \sim 0.11 (\alpha / 0.1)^{-1/2} (m/ 33)^{-2/3}$ and $\dot{m} 
\sim 0.17 (\alpha / 0.1)^{-1/2} (m/ 33)^{-2/3}$ are shown in the {\it 
left and right panels} by long-dashed and dotted lines respectively.
\label{qpo_dip_1}}
\end{figure}

The {\bf \em second chapter} contains the analysis based on the RXTE 
observations of GRS 1915+105 in the flaring state, when the spectrum 
and fast variability of the source were changing on the time scales from 
a few seconds up to $\sim 1000$ seconds (Fig. \ref{grs1915_2_lc_pds}). 
We show that rich variety of source 
bursting behavior can be reduced to a sequence of two types of transitions 
between soft and hard states qualitatively distinguished by their spectral 
and temporal properties. The quasi--periodic oscillations (QPOs) with the 
frequency varying between $\sim$2 and 10 Hz are associated with the 
episodes of harder source spectrum. In each observation we found tight 
correlation between the duration of the hard episode $t_{\rm hard}$ and 
the characteristic QPO frequency $f_{\rm QPO}$. For a half of the 
observations this correlation $t_{\rm hard} \propto f_{\rm QPO}^{-7/3}$ 
matches the relation between the viscous time scale, $t_{\rm visc}$ 
and the Keplerian frequency, $f_{\rm K}$, when both quantities are 
evaluated for various radii in the radiation pressure dominated accretion 
disk (Fig. \ref{qpo_dip_1}). Assuming that the QPO frequency is 
proportional to the Keplerian frequency at the boundary between an 
optically thick accretion disk and a hot comptonization region, the changes 
of the QPO frequency can then be understood as due to variations of this 
boundary position on the viscous time scales. Aforementioned model allows 
to estimate typical value of the disk accretion rate, $\dot{m}$ in GRS 
1915+105 during the flaring state: $\dot{m} \sim (0.1 - 0.2) (\alpha /0.1)^
{-1/2} (M/33 M_{\odot})^{-2/3}$, where $\alpha$ -- viscosity parameter and 
$M$ -- compact object mass. 

In the {\bf \em third chapter} we present the interpretation of the 
spectral and temporal properties of GRS 1915+105 in the framework of 
two--component model of the accretion flow involving hot comptonization 
region near the compact object surrounded by the optically thick accretion 
disk. 

The {\bf fourth part} demonstrates that the comparison of the hard 
X-ray transient detection rates by GRANAT/SIGMA and currently operating 
all--sky monitors (CGRO/BATSE, GRANAT/WATCH) leads to following conclusions 
regarding the spatial and luminosity distributions of the hard X-ray 
transients: 

\noindent (i) the spatial distribution of X-ray Novae exhibits a 
concentration towards the Galactic Center similar to the distribution 
of the visible matter in the Galaxy. In particular, it is more 
concentrated than Galactic disk population.

\noindent (ii) if hard X-ray transients follow distribution of visible 
mass in the Galaxy then their peak luminosity in the $35 - 150$ keV band 
is close to $10^{37}$ {\it erg s$^{-1}$} with a relatively small scatter.

In the {\bf fifth part} the results of application of the bulk motion
comptonization model to the approximation of broad--band spectra of 
galactic black hole candidates in the {\em high/very high} state are 
presented. It is shown that this model gives reasonable description of 
the observational data in the $3.5 - 150$ keV energy domain.

\begin{figure}[htb]
\hbox{
\begin{minipage}{9cm}
\epsfxsize=9.0cm
\epsffile{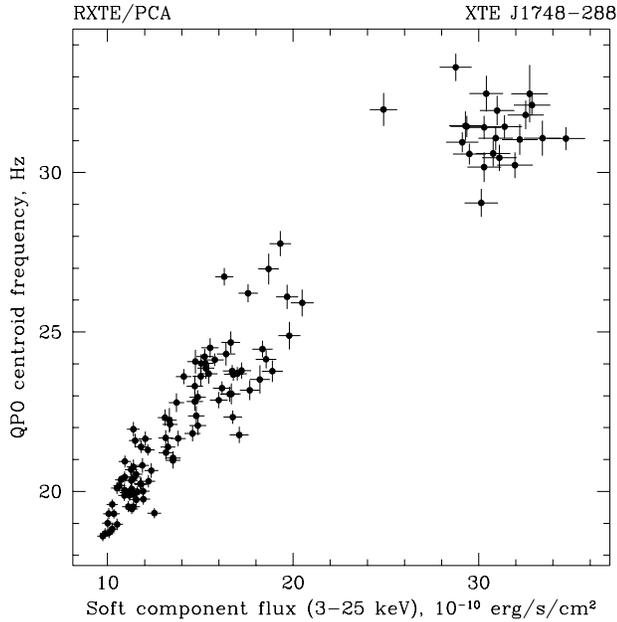}
\end{minipage}
\begin{minipage}{4.5cm}
\caption{\small The dependence of the QPO centroid frequency on the 
X-ray flux in the soft spectral component in the $3 - 25$ keV energy range
(without correction for an interstellar absorption). The parameters of
the soft component are for the best-fit approximation by 
multicolor disk black body plus power law model. PCA data 
for the {\em very high} state observations (June $4 - 11, 1998$) have 
been used. Each point corresponds to the data averaged over 256-s time 
intervals. \label{xte1748_qpo_lum_soft}}
\end{minipage}
}
\end{figure}

\begin{figure}[htb]
\hbox{
\begin{minipage}{9.0cm}
\epsfxsize=9.0cm
\epsffile{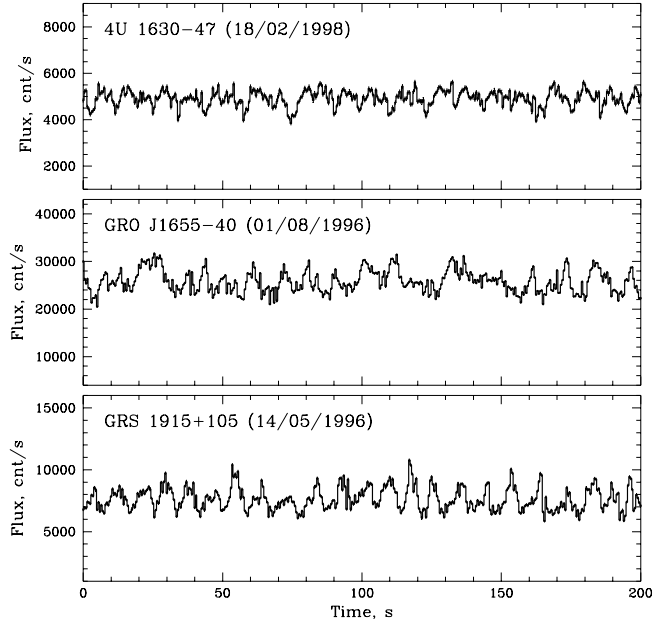}
\end{minipage}
\begin{minipage}{4.5cm}
\caption{\small Similar X-ray variability patterns for 4U1630-47 
({\em upper panel})(Feb. 18, 1998 observation), GRO~J1655--40 
({\em middle panel}) and GRS~1915+105 ({\em lower panel}) according 
to the RXTE/PCA observations ($2 - 30$ keV energy range, fluxes 
correspond to the 5 Proportional Counter Units). \label{4u1630_dip_lc_similar}}
\end{minipage}
}
\end{figure}

\begin{figure}[htb]
\epsfxsize=14.0cm
\epsffile{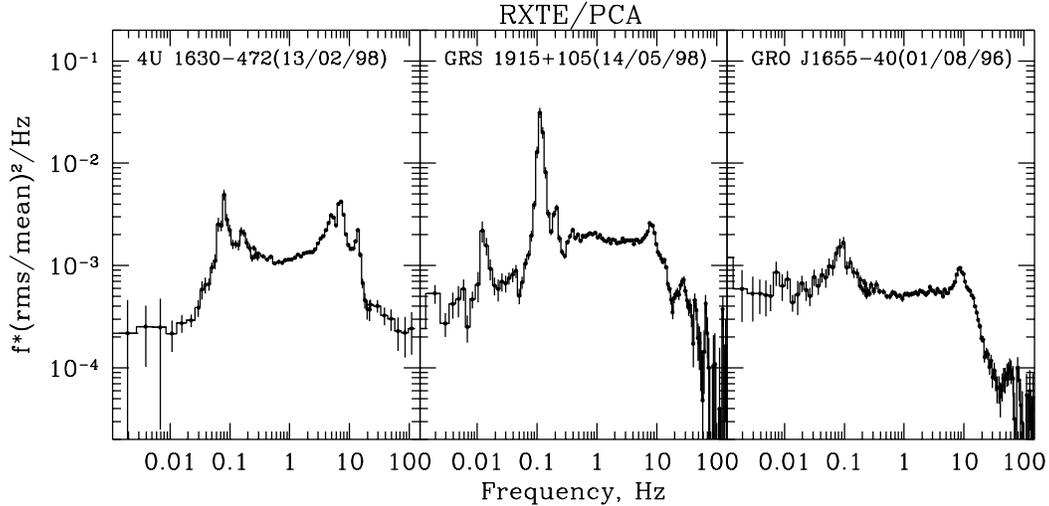}
\caption{\small Power density spectra of 4U~1630--47, GRO~J1655--40 and
GRS~1915+105 for the observations with similar slow variability pattern 
(Fig. \ref{4u1630_dip_lc_similar}) PCA data. \label{4u1630_dip_pds_all}}
\end{figure}

\begin{figure}
\hbox{
\begin{minipage}{9.0cm}
\epsfxsize=9.0cm
\epsffile{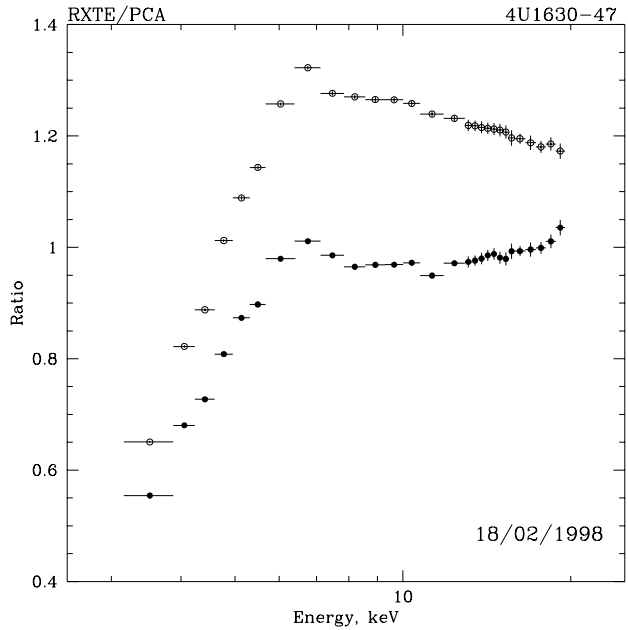}
\end{minipage}
\begin{minipage}{4.5cm}
\caption{\small The changes in 4U1630-47 energy spectrum between the low 
({\em filled circles}) and high ({\em open circles}) flux episodes for the
same observation. Ratios of energy spectra to a power law with photon 
index $\alpha = 2.5$ are shown. The differences in both the temperature 
of soft component and the slope of hard component can be interpreted as 
an inward motion of the accretion disk inner edge during the periods of 
higher flux. \label{4u_high_low_spec_rat}} 
\end{minipage}
}
\end{figure}

\begin{figure}
\hbox{
\epsfxsize=7.0cm
\epsffile{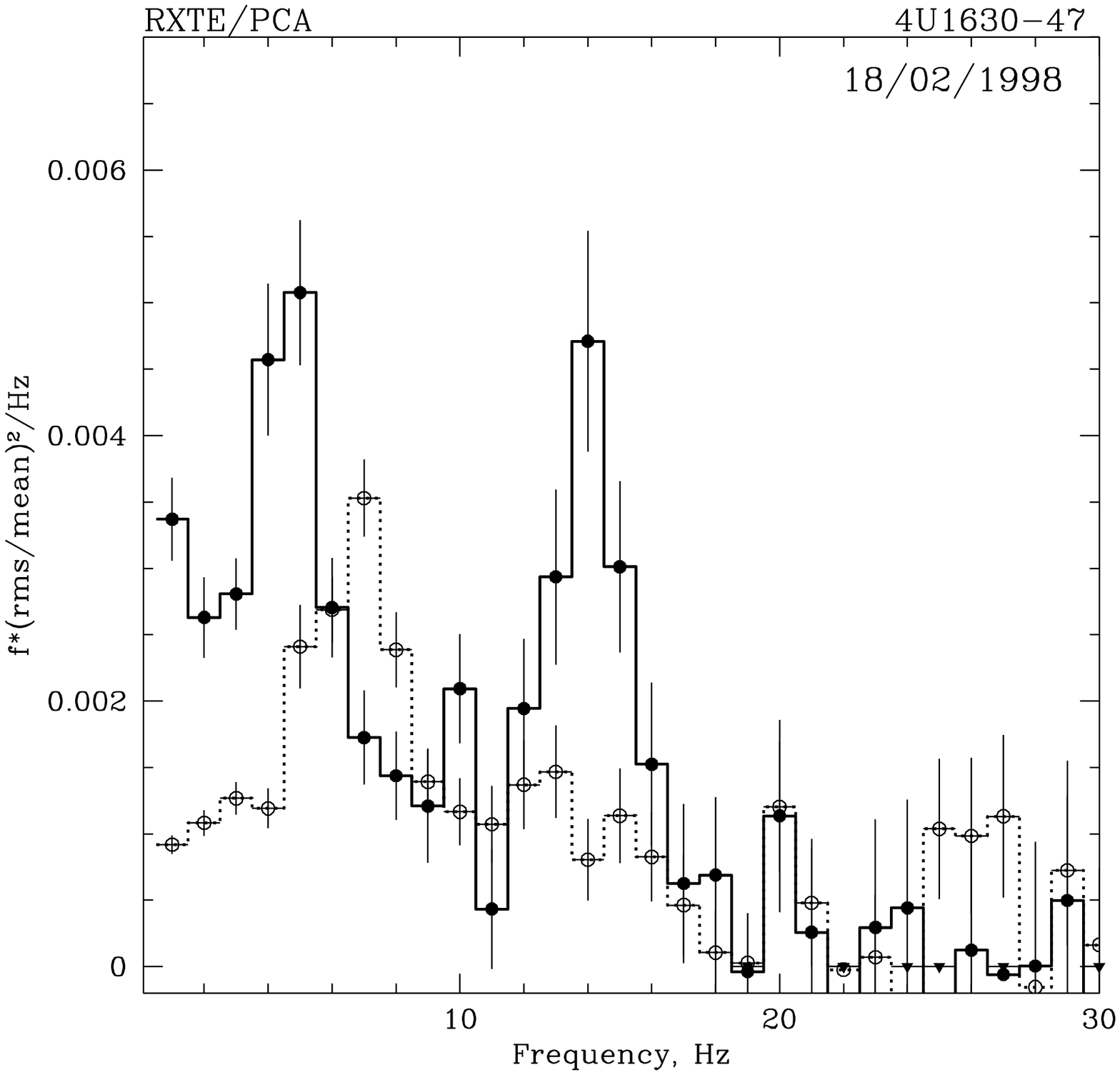}
\epsfxsize=7.0cm
\epsffile{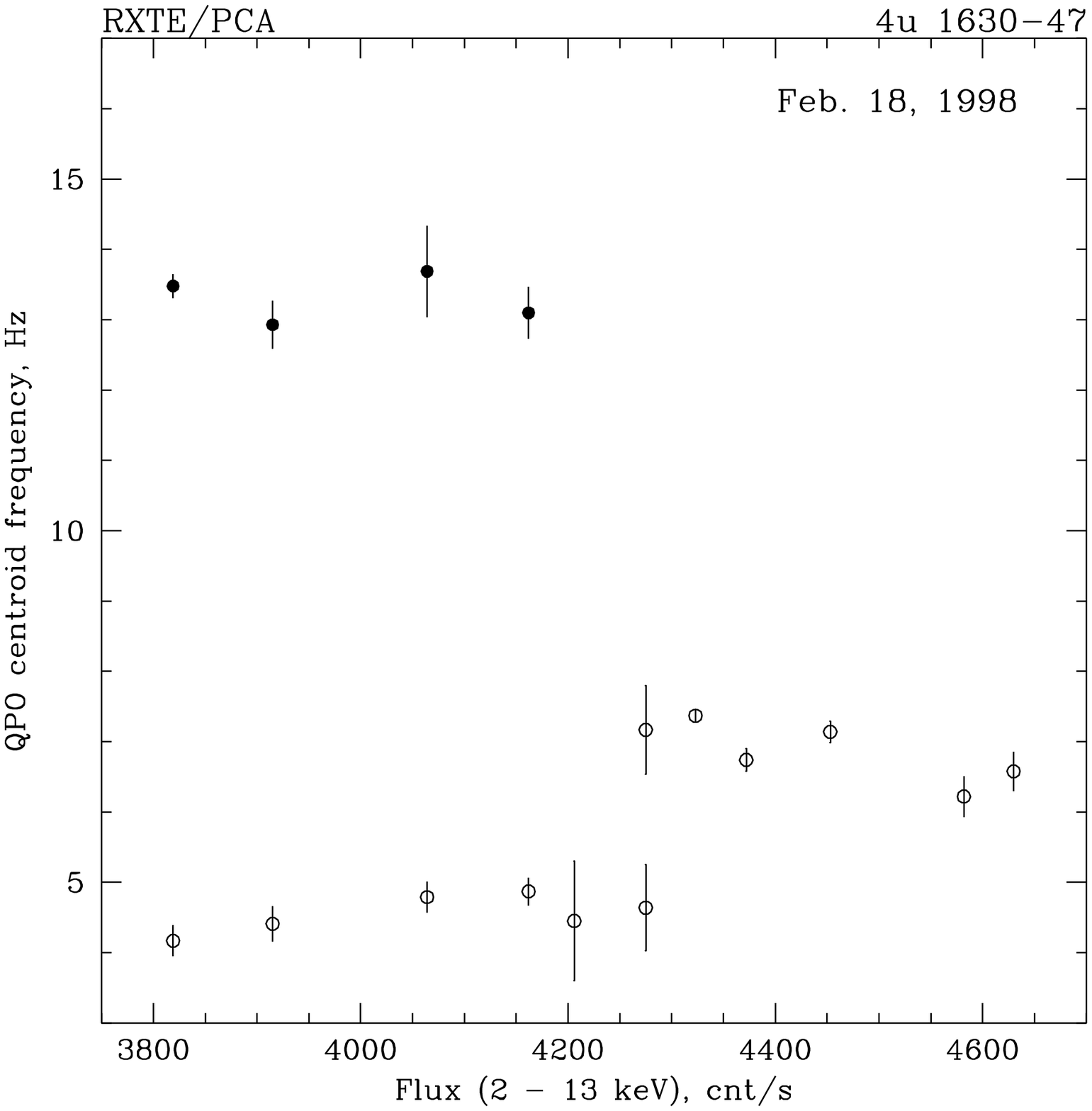}
}
\caption{\small {\bf Left:} Comparison of the source power density spectra 
in units of ${\rm frequency \times (rms/mean)^{2}}$/Hz during the low 
({\em filled circles, solid lines}) and high ({\em hollow circles, dotted 
lines}) flux episodes for the same observation. QPO at $\sim$ 13 Hz is 
prominent for low fluxes, and disappears for high fluxes. This may be 
attributed to the breakthrough of the shock, or to inward motion of the 
accretion disk. {\bf Right:} Dependence of the QPO centroid frequencies 
on the source flux in the $2 - 13$ keV energy band (5 PCUs of PCA detector) 
for the Feb. 18, 1998 observation. The $\sim 13$--Hz QPO feature is not 
detectable for fluxes above $\sim 4200$ cnt/s. The centroid of the second 
QPO feature is jump--shifted to higher frequencies at the same flux level. 
These data may be interpreted as an existence of two distinct quasi-stable 
states with sharp transition between them. \label{4u_high_low_pds}}
\end{figure}

The {\bf sixth part} of the thesis consists of 5 chapters, containing 
the results of systematic analysis of emission properties of several 
Galactic black hole candidate X-ray Novae: KS/GRS 1730--312, 
GRS 1739--278, GRS 1737--31, XTE J1748--288 and 4U 1630--47. Detection 
of the correlated evolution of spectral and timing characteristics of 
these sources is one of the most important results. For example, in the 
case of XTE J1748--288 the correlation between the QPO frequency and 
soft spectral component flux was clearly detected (Fig. 
\ref{xte1748_qpo_lum_soft}). Similarly to the Galactic microquasar 
GRS 1915+105 (Fig. \ref{qpo_soft_lum}), this type of correlation holds 
on a wide range of timescales. Another interesting feature is the 
quasi--regular modulation with periods of $\sim 10 - 20$ s detected 
during the maximum phase of the outburst of 4U 1630--47 (Fig. 
\ref{4u1630_dip_lc_similar}). It is notable 
that somewhat similar type of variability was observed in Galactic 
microquasars GRS~1915+105 and GRO~J1655-40 (Fig. \ref{4u1630_dip_lc_similar}
). We speculate that this mode might be common for black hole binaries 
emitting at certain luminosity level. It is remarkable that in all cases, 
this mode corresponds to a relatively narrow interval of source 
luminosities. Our analysis revealed significant differences in spectral 
and temporal behavior of the source at high and low fluxes during this 
period of time (Fig. \ref{4u_high_low_spec_rat}, \ref{4u_high_low_pds}). 

As in the case of GRS 1915+105 the correlated spectral and timing behavior 
in course of the X-ray Novae outburst can be generally understood in the 
framework of the two-phase model of the accretion flow around the compact 
object (i.e. the composition of a hot inner comptonization region and 
surrounding optically thick accretion disk). The interaction between 
these two distinct regions determines the properties of the spectrum and 
variability of the source. Assuming the QPO phenomenon to be related to 
the dynamical time scale on the boundary between the hot inner region 
and the outer accretion disk, we can treat the observed increase of the 
QPO centroid frequency as an indication of the inward motion of this 
boundary during the rise phase of the outburst. This interpretation is 
supported by simultaneous shift of the maximum of the PDS band-limited 
noise component (depending on the characteristic radius of the inner 
comptonization region) and softening of the energy spectrum. Similarity 
of the rise phases of X-ray Novae outbursts and the outbursts allows 
us to suggest that in all these cases the approach of the inner edge of 
an optically thick accretion disk to the black hole causes the rise of the 
soft X-ray emission and softening of the broad--band energy spectrum. 
Alternatively, the hardening of the energy spectrum and decrease of the 
QPO centroid frequency and frequency of the BLN component maximum observed 
during the final stages of the X-ray Novae outbursts, could be explained 
in terms of the outward motion of the boundary between the inner region 
and an optically thick accretion disk.

The variations in the characteristic QPO frequencies hint at a change 
of some typical radius of the system, in particular, the inner radius 
of the optically thick accretion disk. Because the luminosity of the disk 
depends on both the accretion rate and the inner radius of the disk, we 
can not simply attribute any changes in the source luminosity to the 
change of the accretion rate onto the central object. In fact, the 
luminosity can vary significantly for the same accretion rate, if the 
geometry of the system and relative contribution from disk and corona 
component change. The observation of several plateaus at the peak of 
the outburst of some sources (e.g 4U 1630--47) shows the existence of 
quasi--stationary modes of the accretion for this state.

Part of the research has made use of data obtained through the HEASARC 
Online Service, provided by the NASA/GSFC. 

\end{document}